\documentclass[aps,prl,twocolumn,superscriptaddress,amsmath,amssymb]{revtex4-1}
\usepackage{graphicx}  
\usepackage{dcolumn}   
\usepackage{bm}        
\usepackage{verbatim}   
\usepackage{epstopdf}
\usepackage{xcolor}

\def\CO2{CO$_2$}

\newcommand{\perm}{k}

\makeatletter
\newenvironment{nbytwosubequations}{%
  \refstepcounter{equation}%
  \protected@edef\theparentequation{\theequation}%
  \setcounter{parentequation}{\value{equation}}%
  \setcounter{equation}{0}%
  \def\theequation{%
    \theparentequation\alph{equation}%
    \addtocounter{equation}{1},\alph{equation}%
  }%
  \ignorespaces
}{%
  \setcounter{equation}{\value{parentequation}}%
  \ignorespacesafterend
}
\makeatother

\begin{document}

\title{Convection in porous media with dispersion}

 \author{Baole Wen}
 \email[ ]{wenbaole@gmail.com or baole@ices.utexas.edu}
 \affiliation{Institute of Computational Engineering and Sciences, The University of Texas at Austin, Austin, TX 78712 USA}
 \affiliation{Department of Geological Sciences, Jackson School of Geosciences, The University of Texas at Austin, Austin, TX 78712 USA}
 \author{Kyung Won Chang}
  \email[ ]{Present address: Geomechanics Department, Sandia National Laboratories, Albuquerque, NM 87123 USA}
  \affiliation{Department of Geological Sciences, Jackson School of Geosciences, The University of Texas at Austin, Austin, TX 78712 USA}
 \author{Marc A. Hesse}
 \email[ ]{mhesse@jsg.utexas.edu }
 \affiliation{Institute of Computational Engineering and Sciences, The University of Texas at Austin, Austin, TX 78712 USA}
 \affiliation{Department of Geological Sciences, Jackson School of Geosciences, The University of Texas at Austin, Austin, TX 78712 USA}

\begin{abstract}
We investigate the effect of dispersion on convection in porous media by performing direct numerical simulations (DNS) in a two-dimensional Rayleigh-Darcy domain.  Scaling analysis of the governing equations shows that the dynamics of this system are not only controlled by the classical Rayleigh-Darcy number based on molecular diffusion, $Ra_m$, and the domain aspect ratio, but also controlled by two other dimensionless parameters: the dispersive Rayleigh number $Ra_d = H/\alpha_t$ and the dispersivity ratio $r = \alpha_l/\alpha_t$, where $H$ is the domain height, $\alpha_t$ and $\alpha_l$ are the transverse and longitudinal dispersivities, respectively.  For $\Delta = Ra_d/Ra_m > O(1)$, the influence from the mechanical dispersion is minor; for $\Delta \ll 1$, however, the flow pattern is controlled by $Ra_d$ while the convective flux is $F\sim Ra_m$ for large $Ra_m$, but with a prefactor that has a non-monotonic dependence on $Ra_d$.  Our DNS results also show that the increase of mechanical dispersion, i.e. decreasing $Ra_d$, will coarsen the convective pattern by increasing the plume spacing.  Moreover, the inherent anisotropy of mechanical dispersion breaks the columnar structure of the mega-plumes at large $Ra_m$, if $Ra_d < 5000$. This results in a fan-flow geometry that reduces the convective flux.
\end{abstract}


\maketitle

Convection in porous media controls mass and energy transfer in many natural and engineered applications \citep{Horton1945,Lapwood1948,Phillips09,NB2006}. This subject has received renewed interest due to its potential impact on geological carbon dioxide (\CO2) storage, which allows large reductions of \CO2 emissions from fossil fuel-based electricity generation \citep{Orr2009,Michael2010,Szulczewski2012}. After the \CO2 is injected into the deep saline aquifers, it dissolves into the brine and increases the brine density. The dissolution of \CO2 eventually forms a stable stratification and ensures secure long-term storage \citep{Weir1996,Ennis-KingPaterson2005}.

The rate of \CO2 dissolution is limited by mass transfer of dissolved \CO2 away from the \CO2-brine interface. Diffusive mass transport may take millions of years to saturate the brine \cite{Sathaye2014,Akhbari2017,Wen2018JFM}. However, once the diffusive boundary layer of dissolved \CO2 in brine has grown thick enough, it might become unstable and subsequently, convection sets in and forms descending \CO2-rich plumes. This process greatly increases the \CO2 dissolution rate and significantly reduces the leakage risk of buoyant \CO2 into potable aquifers or into the atmosphere \citep{Roberts2011}.

Dynamics of porous media convection can be studied in either a `one-sided' system where convection is driven by a source of buoyancy on only one boundary, e.g. the solutal convection system \citep{Riaz2006,Neufeld2010,Hewitt2013,Szulczewski2013,Slim2014,Shi2018}, or a `two-sided' system where both of top and bottom boundaries actively drive the convection, e.g. the thermal convection system \citep{Graham1992,Otero2004,Hewitt2012,Hewitt2013,Hewitt2014,Wen2015thesis,Wen2015PRE}. These two systems share many common characteristics in convective pattern and transport properties, although dynamics in the former generally evolve over time while there exists a statistically-steady state in the later. In this study, we focus on the two-sided convective system to perform long-time direct numerical simulations (DNS) for reliable statistically results.

In the absence of mechanical dispersion, the flow pattern and transport flux of convection in porous media are generally thought to be controlled by the molecular Rayleigh number, 
\begin{eqnarray}
    Ra_m = \frac{\perm\Delta\rho gH}{\mu\phi D_m}, \label{Ra_m}
\end{eqnarray}
where $\perm$ is the medium permeability, $\Delta\rho$ the density change between the fresh and the saturated water, $g$ the acceleration of gravity, $H$ the domain height, $\mu$ the dynamic viscosity of the fluid, $\phi$ the porosity, and $D_m$ the molecular diffusion coefficient. At large $Ra_m$, convection appears in the form of columnar plumes fed continually with a series of proto-plumes generated from the diffusive boundary layer. As $Ra_m$ is increased, the inter-plume spacing $\delta$ and the flux $F$ in the quasi-steady convective regime follow specific power-law scalings of $Ra_m$, i.e. $\delta \sim Ra_m^{-\alpha}$ with the positive exponent $\alpha \le 0.5$ \citep{Hewitt2012, Hewitt2013, Hewitt2014, Hewitt2017, Wen2015JFM, WenChini2018JFM}, and $F \sim Ra_m^\beta$ with $\beta = 1$ reported by \cite{Otero2004, Pau2010, Hidalgo2012, Hewitt2012, Slim2014, Wen2012, Wen2013, Wen2015JFM, WenChini2018JFM}, although experimental results have been interpreted to suggest lower exponents $0.8 \le \beta < 1$ \cite{Neufeld2010,Backhaus2011}.

However, recent bench-top experiments on solutal convection in porous media show that $Ra_m$ does not control the convective pattern in typical granular media, because mechanical dispersion is the dominant dissipative mechanism \cite{Liang2018}. Mechanical dispersion in porous media is due to non-uniformities in the flow that cause mixing of the solute \citep{DeJosselindeJong1958,Saffman1959,Bachmat1964}.  In an isotropic and homogeneous porous medium, mechanical dispersion is described by two parameters: the longitudinal and transverse dispersivities $\alpha_l$ and $\alpha_t$, respectively \citep{Bear1961, Scheidegger1961, Jong1961}. Therefore, the hydrodynamic dispersion tensor in the fixed Cartesian reference frame can be expressed as
\begin{eqnarray}
    \mathbf{D}^* = D_m\mathbf{I} + (\alpha_l - \alpha_t)\frac{\mathbf{u}^*\mathbf{u}^*}{|\mathbf{u}^*|} + \alpha_t|\mathbf{u}^*|\mathbf{I}, \label{D_dim}
\end{eqnarray}
where $\mathbf{I}$ denotes the identity matrix and the mechanical dispersion scales linearly with the interstitial fluid velocity  $\mathbf{u}^*$. As long as $|\mathbf{u}^*| \ll D_m/\alpha_l$, $\mathbf{D}^* \approx D_m\mathbf{I}$, so that molecular diffusion dominates over hydrodynamic dispersion; when $|\mathbf{u}^*| \gg D_m/\alpha_l$, however, the mechanical dispersion starts to dominate.

Recent studies by \cite{Ghesmat2008, Hidalgo2009, Ghesmat2011, Emami-Meybodi2015, Wang2016, Suekane2017, Liang2018} indicate that hydrodynamic dispersion significantly affects the flow pattern and mass transport of convection in porous media at large $Ra_m$. The numerical simulations by \cite{Hidalgo2009, Ghesmat2011} show that hydrodynamic dispersion enhances the convective mixing and greatly reduces the onset time for convection; however, the experiments by \cite{Wang2016, Liang2018} reveal that the mechanical dispersion coarsens the convective pattern and reduces the increase of convective flux with increasing permeability $k$. Particularly, the systematic experiments by \cite{Liang2018} illustrate that adjusting $Ra_m$ via changing the density difference $\Delta\rho$ or the medium permeability $k$ may result in distinct convective characteristics due to hydrodynamic dispersion. For fixed $\Delta\rho$, increasing $k$ (via choosing a larger glass bead diameter $d$ as $k \sim d^2$) raises $Ra_m$ but \emph{enlarges} the inter-plume spacing $\delta$; for fixed $k$, however, $\delta$ is nearly fixed for increasing $\Delta\rho$. Secondly, for fixed $\Delta\rho$, the dissolution flux $F$ does not increase linearly with $k$ and is lower than expected at high $k$; for fixed $k$, in contrast, $F\sim cRa_m^1$ with decreasing prefactor $c$ as $k$ is increased. Nevertheless, the vertical velocity, via measuring the speed of the fastest descending fingertip, approximately increases linearly with both $\Delta\rho$ and $k$. Some of the above findings contradict the classical predictions made in the absence of mechanical dispersion. 

To understand the effect of dispersion on convection, we perform DNS in a two-dimensional (2D), rectangular, homogeneous and isotropic Rayleigh-Darcy domain.  We aim to identify the dimensionless parameters governing convection in porous media with hydrodynamic dispersion, determine the scaling law for the quasi-steady convective flux, and quantify the contribution of molecular diffusion and mechanical dispersion to the hydrodynamic dissipation. As mentioned earlier, we focus on a two-sided convective system for long-time statistical results. 

In previous studies, the dispersivity, $\alpha_l$ or $\alpha_t$, and the molecular diffusivity $D_m$ are combined to define the characteristic length and time scales or the Rayleigh number \citep{Hidalgo2009,Ghesmat2011}. In this work, however, we rescale the system using the domain height $H$, the buoyancy velocity $U = \perm\Delta\rho g/(\mu\phi)$, and the convective timescale $T_c = H/U$. We show below that this allows us to decouple the parameters controlling the flow pattern and the flux which simplifies the discussion. Based on these scales, we obtain the dimensionless equations
\begin{subequations}\label{eq:gov_pde_non_dim}
\begin{eqnarray}
& \dfrac{\partial C}{\partial t} + \mathbf{u}\cdot\nabla C = \nabla\cdot(\mathbf{D} \nabla C), \label{Solute_nondim}\\
& \mathbf{u} = -{\nabla}p - C{\bf e}_{z}, \label{Darcy_nondim} \\
&\nabla\cdot\mathbf{u} = 0, \label{Continuity_nondim}
\end{eqnarray}
\end{subequations}
where $C$, $\mathbf{u} = (u,w)$, and $p$ are the dimensionless forms of concentration, velocity, and pressure, respectively, and ${\bf e}_{z}$ is a unit vector in $z$ (upward) direction. The dimensionless hydrodynamic dispersion tensor is then given by
\begin{eqnarray}
    \mathbf{D} = Ra_m^{-1}\mathbf{I} + Ra_d^{-1}\left[(r - 1)\frac{\mathbf{u}\mathbf{u}}{|\mathbf{u}|} + |\mathbf{u}|\mathbf{I}\right], \label{D_nondim}
\end{eqnarray}
and characterized by the molecular Rayleigh number $Ra_m$ defined in (\ref{Ra_m}) and two additional parameters,
\begin{nbytwosubequations}
\begin{eqnarray}
    Ra_d = \frac{H}{\alpha_t} \quad \mbox{and} \quad r=\frac{\alpha_l}{\alpha_t} \label{Rad_r}, 
\end{eqnarray}
\end{nbytwosubequations}
which are referred to as dispersive Rayleigh number and dispersivity ratio, respectively.  The dissipation by mechanical dispersion increases with decreasing $Ra_d$.

The flow is assumed to be periodic laterally with a impermeable top and bottom boundaries.  Solute concentration along the top  and bottom boundaries is unity and null, respectively. Hence, the boundary conditions at the top and the bottom are given by 
\begin{eqnarray}
	\left.C\right|_{z=1} =  1 \; \mbox{and} \; \left.w\right|_{z=1} = 0; \; \left.C\right|_{z=0} =  \left.w\right|_{z=0} = 0. \label{BCs_nondim}
\end{eqnarray}
Note that the problem posed by (\ref{eq:gov_pde_non_dim}) and (\ref{BCs_nondim}) is formally identical to the two-sided thermal convection problem in which the domain is heated from below and cooled from above. Here, (\ref{eq:gov_pde_non_dim}) and (\ref{BCs_nondim}) are solved numerically using a Fourier-Chebyshev-tau pseudospectral solver developed in \cite{Wen2015JFM, WenChini2018JFM}, the temporal discretization is achieved using a three-step semi-implicit Runge-Kutta scheme \citep{Nikitin2006}, and the numerical scheme is parallelized using the Message Passing Interface (MPI). In order to obtain reliable statistical results, the DNS are performed up to $O(10^3)$ convective time units.  As the transverse dispersivity is usually an order of magnitude less than the longitudinal dispersivity \citep{Gelhar1992,Johannsen2002,Muniruzzaman2017}, we set $r = 10$ in all simulations.

To quantify the flow, we measure the convective flux $F$ at the top wall, 
\begin{eqnarray}
    F = \left. \left\langle\overline{\frac{\partial C}{\partial z}} + \frac{Ra_m}{Ra_d}\overline{|u|\frac{\partial C}{\partial z}}\right\rangle\right|_{z=1} = F_m + F_d, \label{flux}
\end{eqnarray}
where the angle bracket and the overbar denote the long-time and the horizontal averages, respectively, the first term on the right side of (\ref{flux}) represents the flux at the boundary via pure molecular diffusion $F_m$, and the second term represents the flux via mechanical dispersion $F_d$.  We also measure the inter-plume spacing $\delta$ by time-averaging the dominant Fourier mode number at the interior, the mean horizontal velocity at the top wall, $\tilde{u} = \langle\overline{|u|}\rangle|_{z=1}$, the mean vertical velocity at the interior, $\tilde{w} = \langle\overline{|w|}\rangle|_{z=\frac{1}{2}}$, and the magnitude of the time-averaged $w$ extremum value at the interior, $w_m = \langle\max(|w|_{z=\frac{1}{2}})\rangle$.

\begin{figure}[t!]
    {\centering
    \includegraphics[width=3.4in]{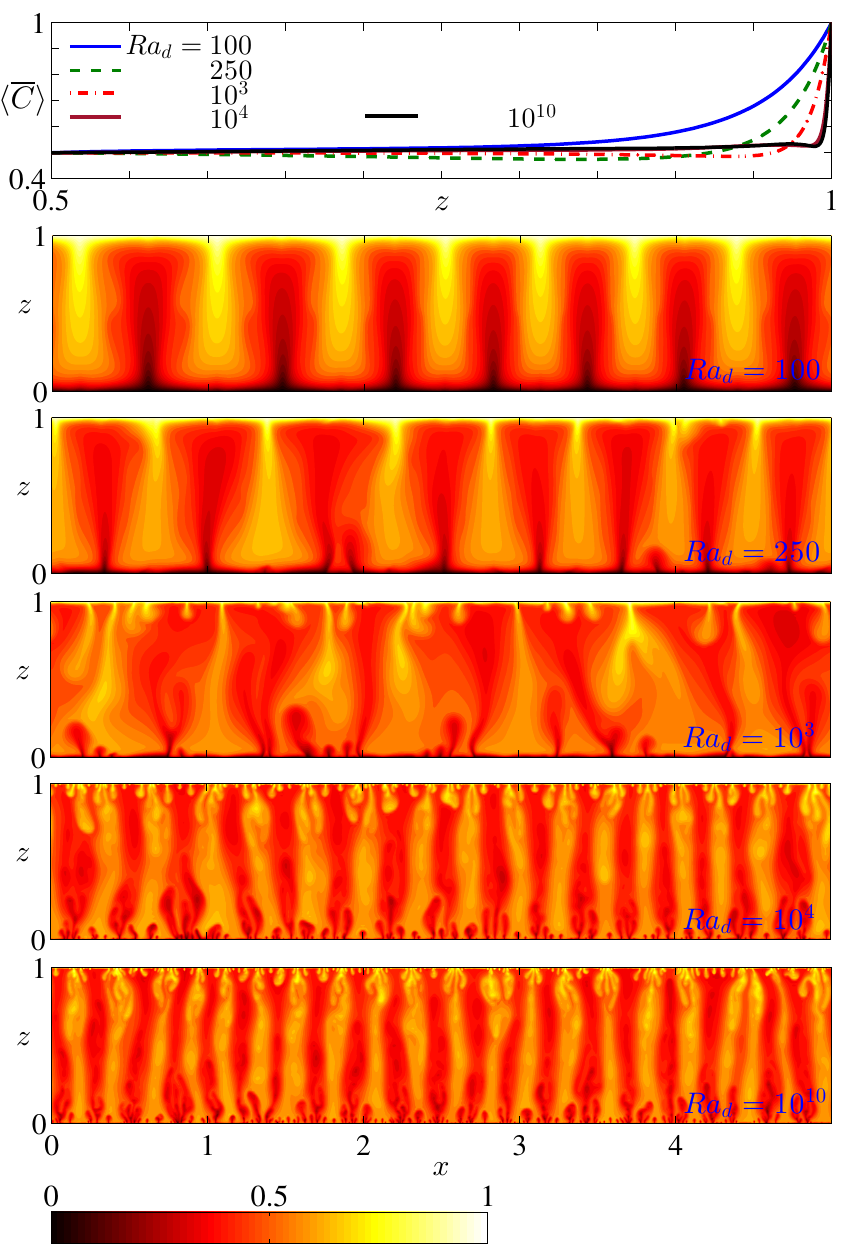}
    }
  \caption{Time-averaged horizontal-mean concentration profile $\langle\overline{C}\rangle$ and snapshots of the concentration field $C$ from DNS at $Ra_m = 20000$ and $r=10$ for different $Ra_d$. The domain aspect ratio is $L=5$.}  \label{Pattern_Ram20000}
\end{figure}

\begin{figure}[t!]
    {\centering
    \includegraphics[width=3.4in]{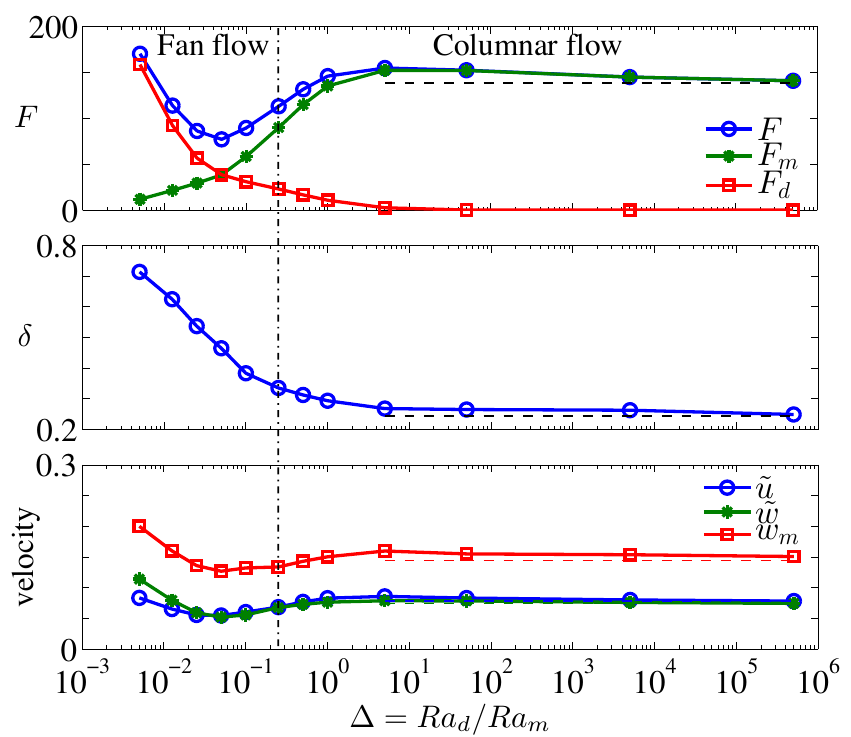}
    }
  \caption{Statistical DNS results of convection  at $Ra_m = 20000$ and $r=10$ for different $Ra_d$. The domain aspect ratio is $L=5$. The dashed lines denote the results in the absence of mechanical dispersion and the dashed-dot line separates the fan-flow and the columnar-flow regions.}  \label{F_delta_u_w_Ram19905}
\end{figure}

Figures~\ref{Pattern_Ram20000} and \ref{F_delta_u_w_Ram19905} show the variation of the convective flow pattern and the corresponding statistical DNS results as a function of $Ra_d$ for $Ra_m=20000$. When the smallest diffusive length scale $1/Ra_m$ is much larger than the pore scale of the medium $d/H$, i.e. $Ra_d \gg r  Ra_m$ as $\alpha_t\approx d/r$ \cite{Saffman1959,Oswald2004}, the molecular diffusion dominates the hydrodynamics dispersion \citep{Hewitt2012,Slim2014,Liang2018}. Our DNS results reveal that only for $\Delta \equiv Ra_d/Ra_m \gtrsim 10^5$, the mechanical dispersion can be completely negligible so that the convection converges to the classical columnar flow \cite{Otero2004, Hewitt2012, Wen2015JFM}. When $O(1) < \Delta < 10^5$, the relatively weak mechanical dispersion slightly enhances the convective transport but the flow still retains the columnar structure. For $\Delta < O(1)$, however, the mechanical dispersion starts to apparently affect the convective pattern and flux: the convection transitions to a fan flow with laterally expanding mega-plumes along the vertical flow direction, and the convective flux is reduced to approximately 50\% of the high-$Ra_d$ value at $\Delta = 0.05$. 

Increasing dispersion thickens the diffusive boundary layer, smooths the small-scale plumes near the walls, and stabilizes the flow. Eventually, the convection becomes steady at $Ra_d = 100$ (Fig.~\ref{Pattern_Ram20000}) and the flux is again increased for $\Delta \le 0.05$ due to the large magnitude of the effective diffusion coefficient induced by the mechanical dispersion. Moreover, it is also seen from Fig.~\ref{F_delta_u_w_Ram19905} that the hydrodynamic dispersion coarsens the flow pattern by increasing $\delta$, and the mean buoyancy velocities at the top and the interior, $\tilde{u}$, $\tilde{w}$ and $w_m$, roughly follow the same trend with the convective flux as a function of $Ra_d$. It should be noted that the $w$ extremum value, $w_m$, becomes nearly constant for $0.025 \le \Delta \le 0.25$.

\begin{figure}
    {\centering
    \includegraphics[width=3.4in]{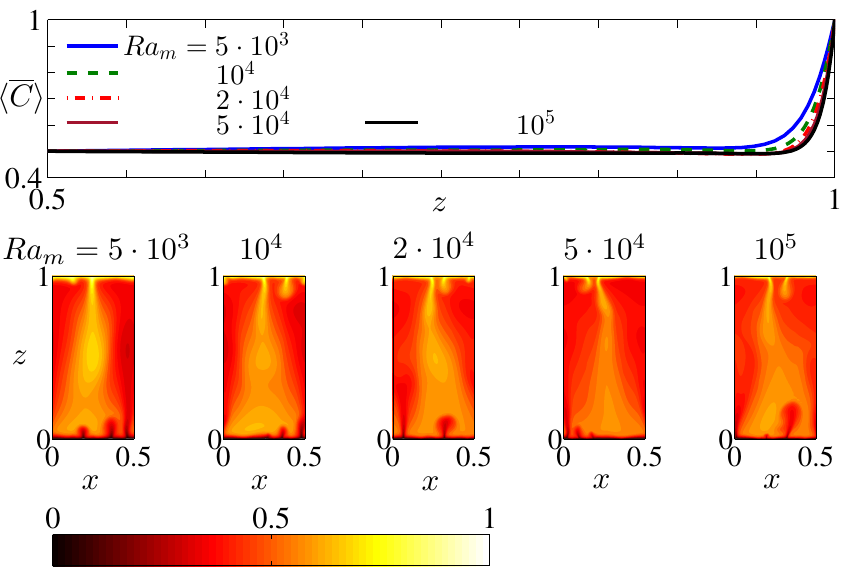}
    }
  \caption{Time-averaged horizontal-mean concentration profile $\langle\overline{C}\rangle$ and snapshots of the concentration field $C$ from DNS at $Ra_d = 1000$ and $r=10$ for different $Ra_m$. For $Ra_m\le20000$, the domain aspect ratio is $L=5$; while for $Ra_m>20000$, DNS are performed in a small unit $L=0.5$ where there only exists a single rising and descending mega-plume but the turbulent convection still sustains itself.}  \label{Pattern_Rad1000}
\end{figure}

\begin{figure}
    {\centering
    \includegraphics[width=3.4in]{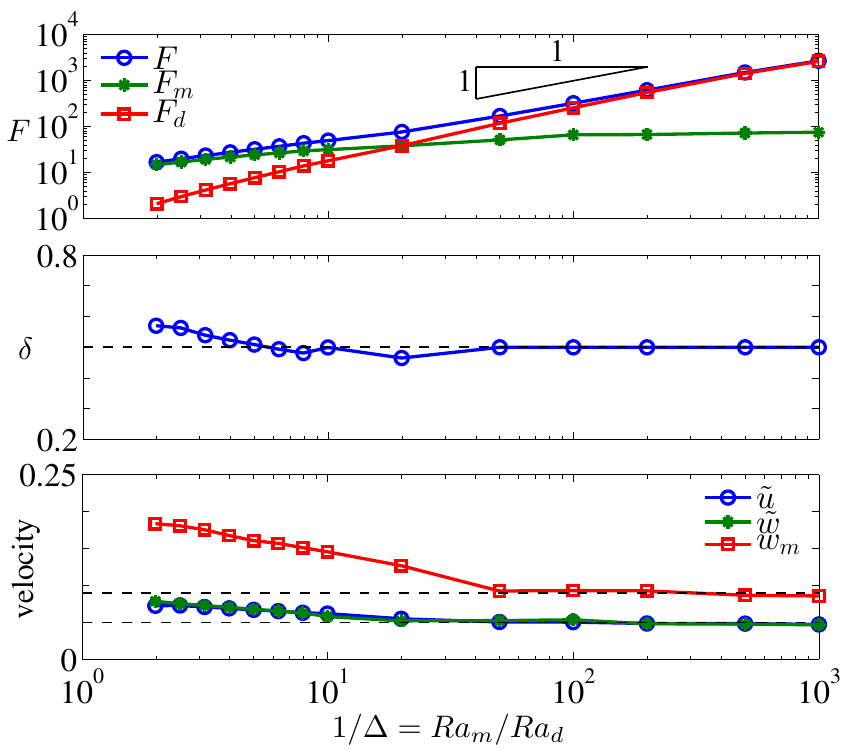}
    }
  \caption{Statistical DNS results of convection at $Ra_d = 1000$ and $r=10$ for different $Ra_m$. $L$ is as in Fig.~\ref{Pattern_Rad1000}.}  \label{F_delta_u_w_Rad1000}
\end{figure}

Figures~\ref{Pattern_Rad1000} and \ref{F_delta_u_w_Rad1000} show the convective pattern and the corresponding statistical DNS results as a function of $Ra_m$ for $Ra_d=1000$. The convection basically remains a fan-flow structure at $Ra_d=1000$ as $Ra_m \rightarrow \infty$. In particular, the inter-plume spacing $\delta$ is nearly invariant when $\Delta \lesssim 0.2$; the mean velocities $\tilde{u}$ and $\tilde{w}$ are roughly unchanged after $\Delta \lesssim 0.05$; and the horizontal-mean concentration profile $\overline{C}$ becomes almost fixed for $\Delta \lesssim 0.02$, so that at the top and the bottom, the flux due to molecular diffusion (i.e. $F_m$) levels off. In short, at sufficiently large $Ra_m$, the flow pattern and the statistical system quantities (i.e. $\overline{C}$, $\delta$, $\tilde{u}$, $\tilde{w}$ and $w_m$) are independent of $Ra_m$. Actually, as $Ra_m \rightarrow \infty$, the hydrodynamic dispersion tensor~(\ref{D_nondim}) reduces to
\begin{eqnarray}
    \mathbf{D} \rightarrow Ra_d^{-1}\left[(r - 1)\frac{\mathbf{u}\mathbf{u}}{|\mathbf{u}|} + |\mathbf{u}|\mathbf{I}\right], \label{D_nondim_inftyRam}
\end{eqnarray}
so that $Ra_d$ becomes the only parameter controlling the dynamics of the system (for fixed $r$). Thus, at large $Ra_m$ the concentration field $C$ and the buoyancy velocity $\mathbf{u}$ are determined by the dispersive Rayleigh number $Ra_d$, as confirmed by our DNS data. Once $C$ and $\mathbf{u}$ become invariant in the limit of $Ra_m \rightarrow \infty$, $F_m \sim c_1$ and $F_d \sim c_2\cdot Ra_m^1$ with the constants $c_1$ and $c_2$ determined by $Ra_d$. 

A natural question concerns how the mechanical dispersion affects convection in the $(Ra_m,Ra_d)$ parameter space. For $\Delta > O(1)$, the influence from the mechanical dispersion is minor, so that both the convective pattern and flux are controlled by $Ra_m$; for $0.02 \lesssim \Delta < O(1)$, both the molecular diffusion and the mechanical dispersion are important to convection, e.g. they equally affect the flux at $\Delta \approx 0.05$; for $\Delta < 0.02$, the mechanical dispersion dominates the hydrodynamic dispersion: the flow pattern is determined by $Ra_d$, e.g. ${C} = {C}(Ra_d)$, $\delta = \delta(Ra_d)$ and $\mathbf{u} = \mathbf{u}(Ra_d)$, while the flux is predominantly controlled by $Ra_m$, e.g. $F \sim c_2(Ra_d)\cdot Ra_m^1$. 

Below we apply these results to the recent laboratory experiments on solutal convection in porous media \cite{Liang2018}. In granular media, the mechanical dispersion is proportional to grain size, $\alpha_l\sim d$, so that the appropriate dispersive Rayleigh number is $Ra_d\approx rH/d$ \cite{Saffman1959,Oswald2004}. Hence, increasing the grain size from 0.8 mm to 4 mm simultaneously increases $Ra_m$ from $1.4\cdot 10^4$ to $5.0\cdot 10^5$ ($\Delta\rho = 9.3$ kg/m$^3$) but decreases $Ra_d$ from 3750 to 750, thereby reducing $\Delta$ from 0.3 to 1.5$\cdot10^{-3}$. For $\Delta < 0.02$, the mechanical dispersion dominates the hydrodynamic dispersion and determines the convective pattern.
Therefore, increasing the grain size at fixed $\Delta\rho$ intensifies the mechanical dispersion and coarsens the convective pattern. On the other hand, varying $\Delta\rho$ at fixed $k$ only changes $Ra_m$ and does not affect the flow pattern set by $Ra_d$.  

Moreover, for fixed $d$, the prefactor $c_2(Ra_d)$ is constant so that the convective flux, $F \sim c_2\cdot Ra_m^1$, increases linearly with $\Delta\rho$; while for fixed $\Delta\rho$, $F$ is lower than expected at higher $k$ since the pattern transitions from columnar flow to fan flow  as $Ra_d$ declines  (Fig.~\ref{F_delta_u_w_Ram19905}). However, this reduction in $F$ is accompanied only by a slight reduction in $w_m$ (Fig.~\ref{F_delta_u_w_Ram19905}), which is consistent with the experimental observation that the speed of the fastest fingers increases approximately linearly with both $\Delta\rho$ and $k$ \cite{Liang2018}.

\begin{figure}
    {\centering
    \includegraphics[width=3.4in]{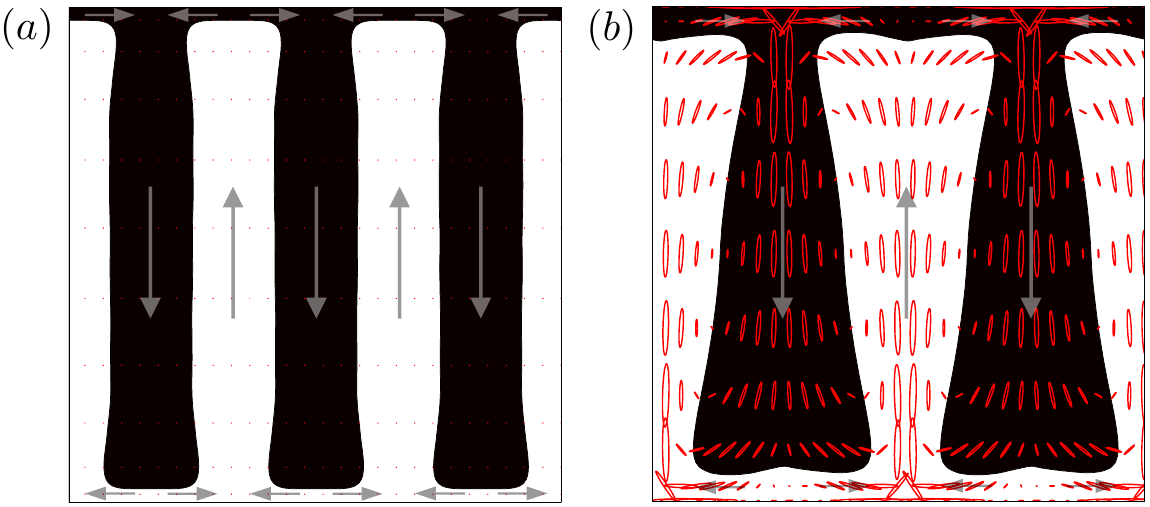}
    }
  \caption{Schematics showing the distribution of the hydrodynamic dispersion tensor in the form of ellipse. ($a$): columnar flow in the absence of mechanical dispersion; ($b$): fan flow with mechanical dispersion. 
  In ($a$), $\mathbf{D}^* = D_m\mathbf{I}$ is homogeneous and isotropic; in ($b$), the anisotropy of the hydrodynamic dispersion leads to an asymmetry between the rising and the descending mega-plumes near the walls.}  \label{Schematics}
\end{figure}

Our DNS results and analysis above reveal that at sufficiently large $Ra_m$, the convective pattern is determined by the dispersive Rayleigh number $Ra_d$: the convection appears in the form of columnar flow at $Ra_d \ge 5000$ and then transitions to a fan flow at $Ra_d < 5000$. Although the convection also exhibits a fan-flow structure at small and moderate $Ra_m$ in the absence of mechanical dispersion \citep{Graham1992, Otero2004}, the physics are different. The fan-flow structure here is due to the inherent anisotropy of mechanical dispersion. 
As shown in Fig.~\ref{Schematics}, near the top and the bottom walls the flow between the neighboring plumes is dominantly horizontal, so the inter-plume spacing is set by the lateral dispersion $D_{xx}^w \approx D_m + \alpha_lu^w$ and the thickness of the diffusive boundary layer is significantly affected by the vertical dispersion $D_{zz}^w \approx D_m + \alpha_tu^w$. At the roots of the plumes, however, the flow is dominantly vertical, so the plume width is controlled by the lateral dispersion $D_{xx}^r \approx D_m + \alpha_tw^r$. The mass conservation of the incompressible flow requires $u^w \approx w^r$ near the wall. Hence, the inherent anisotropy of the mechanical dispersion, $r \gg 1$, leads to $D_{xx}^w \gg D_{xx}^r$, and thereby the increment of the inter-plume spacing is much larger than that of the plume width. This asymmetry results in the fan-flow structure and reduces the transport efficiency.

\bibliography{PRL,mendeley}

\end{document}